\documentclass[a4paper,12pt]{article}
\newcommand{\C}{\mathbb{C}}
\newcommand{\R}{\mathbb{R}}

\usepackage{amssymb,dsfont,amsmath,graphicx,esint}

\begin{document}

\begin{center}
{\Large \bf Integrable models for shallow water with energy
dependent spectral problems }

\vskip.8cm

{\Large \it Rossen  I. Ivanov  \footnote{E-mail:
Rossen.Ivanov@dit.ie,} and Tony Lyons \footnote{E-mail:
Tony.Lyons@mydit.ie,}  } \vskip.8cm School of Mathematical
Sciences, Dublin Institute of Technology,

Kevin Street, Dublin 8, Ireland

\end{center}

\vskip1.32cm

\begin{abstract}
\noindent We study the inverse problem for the so-called operators
with energy depending potentials. In particular, we study spectral
operators with quadratic dependance on the spectral parameter. The
corresponding hierarchy of integrable equations includes the
Kaup-Bousinesq equation. We formulate the inverse problem as a
Riemann-Hilbert problem with a $\mathbb{Z}_2$ reduction group. The
soliton solutions are explicitly obtained.  \vskip.4cm

\noindent {\bf PACS}: 05.45.Yv, 02.20.Sv

\vskip.4cm

\noindent {\bf Key Words}: Inverse Scattering Method, Nonlinear
Evolution Equations, Solitons.
\end{abstract}
\section{Introduction}
The last decades witnessed an explosion in the complexity and
sophistication of mathematical theories for fluids and in
particular for water waves. The soliton theory has been always at
the center of these developments, such as from its early days the
soliton theory has transformed and enhanced enormously the
mathematical description of nonlinear wave propagation. The
simplest and best known integrable water-wave equations belong to
the Korteweg-de Vries family. For some classical and modern
aspects of the theory of water waves, nonlinear waves and soliton
theory we refer to the following monographs and the references
therein:
\cite{Abl81,AC11,FaTa,gvy2008,HSS2009,J97,N85,PS11,W,ZMNP}.

There are classes of soliton equations whose associated spectral
problems are polynomial in the spectral parameter. They are known
also as soliton equations with 'energy dependent potentials' due
to the analogy with the Schr\"odinger equation in Quantum
Mechanical context, whose spectrum represents the energy levels of
the Quantum Mechanical system. Some of these integrable systems
appear as water waves models, most notably the Kaup-Boussinesq
equation \cite{K75,W,EGP01} and the two-component Camassa-Holm
equation \cite{CI08,HSS2009,HI2011,I2009}. Other systems of this
type are studied e.g. in \cite{AF89,AF91,I2006,BPZ01}.

In what follows we study an integrable system which arises as a
compatibility condition of the following two linear operators (Lax
pair):
\begin{eqnarray}\label{Lax.1}
\Psi_{xx} = \left(-\lambda^{2} + \lambda u(x,t) + \frac{\kappa}{2}u^{2}(x,t) + \eta(x,t)\right)\Psi(x,t;\lambda)
\label{Lax.1}\\
\Psi_{t}= -\left(\lambda
+\frac{1}{2}u(x,t)\right)\psi_{x}(x,t;\lambda) +
\frac{1}{4}u_{x}(x,t)\Psi(x,t;\lambda).
\end{eqnarray}
Here $\kappa$ is an arbitrary constant, while $\lambda$ is the
spectral parameter. The consistency condition
$\psi_{xxt}(x,t;\lambda) = \psi_{txx}(x,t;\lambda)$ produces a
system of equations for the functions $u(x,t)$ and $\eta(x,t)$:
\begin{eqnarray}{\label{KdV.2}}
&\phantom{*}& u_{t}+ \eta_{x} + \left(\frac{3}{2}+\kappa\right)uu_{x}=0,\\
&\phantom{*}& \eta_{t}-
\frac{1}{4}u_{xxx}+(u\eta)_x-\left(\frac{1}{2}+\kappa\right)u\eta_{x}
-\kappa \left(\frac{1}{2}+\kappa\right)u^2u_{x}=0.
\end{eqnarray}
Upon choosing $\kappa = -\frac{1}{2}$ these simplify to the well
known Kaup-Boussinesq (or KB for short) equation:
\begin{eqnarray}{\label{KB}}
&\phantom{*}& u_{t}+ \eta_{x} + uu_{x}=0,\\
&\phantom{*}& \eta_{t}- \frac{1}{4}u_{xxx}+(u\eta)_x=0.
\end{eqnarray}

The KB equation is introduced as a water-wave model in \cite{K75}
where also the inverse scattering is studied for functions with
constant limits at $x\to \pm \infty$. As a water-wave model it
also appears in \cite{W,I2009,EGP01,GP01}, the hierarchy of
Hamiltonian structurs is given in \cite{P01}, specific solutions
are studied in \cite{KKU,EGK05,MY79}. Energy-dependant spectral
problems like (\ref{Lax.1}) are studied also in
\cite{J72,JJ72,JJ76,JJ81,MY79,SS97,Lap07}.

Our aim will be to formulate the inverse scattering as a
Riemann-Hilbert Problem (RHP) in the case when $u(x,t)$ and
$\eta(x,t)$ are real, rapidly decaying functions at $x\to \pm
\infty$, taking into account the underlying reductions and to
obtain the simplest soliton solutions.

\section{The spectral problem}

Introducing an auxiliary function
\begin{equation}w(x,t)=\frac{\kappa}{2}u^{2}(x,t) +
\eta(x,t) \label{w} \end{equation} we consider the following two
'conjugate' spectral problems related to (\ref{Lax.1}):
\begin{equation}\label{Lax.2}
\Psi_{xx}(x,\lambda,\sigma) = \left(-\lambda^{2} + \sigma\lambda
u(x) + w(x)\right)\Psi(x,\lambda,\sigma),
\end{equation}
where $\sigma=\pm 1$. The $t$-dependence will be suppressed where
possible for the sake of simplicity.

We specify that $u(x), w(x)$  as well as $\eta(x)$ belong the the
Schwartz class of functions (the space of rapidly decreasing
functions) $\mathcal{S}(\mathbb{R})$.  It follows from this
requirement, that solutions $\psi_{1}(x,t;\lambda)$ and
$\psi_{2}(x,t;\lambda)$ exist such that,
\begin{equation}\label{basis.plus}
      \begin{array}{c}
        \psi_{1}(x,\lambda,\sigma) \to e^{-i\lambda x} \\
        \psi_{2}(x,\lambda,\sigma) \to e^{+i\lambda x} \\
      \end{array}\Bigg\}, \quad x \to +\infty
\end{equation}
Similarly we define a basis of eigenfunctions for (\ref{Lax.2}) according to
\begin{equation}\label{basis.minus}
      \begin{array}{c}
        \phi_{1}(x,\lambda,\sigma) \to e^{-i\lambda x} \\
        \phi_{2}(x,\lambda,\sigma) \to e^{+i\lambda x} \\
      \end{array}\Bigg\}, \quad x \to -\infty
\end{equation}

These eigenfunctions are called Jost Solutions. Since the Jost
solutions oscillate when $\lambda \in \mathbb{R} $, the real
spectrum fills in the real line.

The bases
$\{\psi_{1}(x,\lambda,\sigma),\psi_{2}(x,\lambda,\sigma)\}$ and
$\{\phi_{1}(x,\lambda,\sigma),\phi_{2}(x,\lambda,\sigma)\}$
constitute independent bases of solutions to (\ref{Lax.2}) and as
such, we may write
\begin{eqnarray}\label{scattering.eqn}
       \left(
                  \begin{array}{c}
                    \phi_{1}(x,\lambda,\sigma) \\
                    \phi_{2}(x,\lambda,\sigma) \\
                  \end{array}
                \right) = \left(
                            \begin{array}{cc}
                              T_{11}(\lambda,\sigma) & T_{12}(\lambda,\sigma) \\
                              T_{21}(\lambda,\sigma) & T_{22}(\lambda,\sigma) \\
                            \end{array}
                          \right)\left(
                                        \begin{array}{c}
                                          \psi_{1}(x,\lambda,\sigma) \\
                                          \psi_{2}(x,\lambda,\sigma) \\
                                        \end{array}
                                      \right).
\end{eqnarray}
The matrix
\begin{equation}\label{scattering.matrix}
    \mathbf{T}(\lambda,\sigma)  = \left(
                    \begin{array}{cc}
                              T_{11}(\lambda,\sigma) & T_{12}(\lambda,\sigma) \\
                              T_{21}(\lambda,\sigma) & T_{22}(\lambda,\sigma) \\
                            \end{array}
                          \right)
\end{equation}
is the scattering matrix for spectral problem (\ref{Lax.2}).

Under the involution $(\lambda,\sigma) \to (-\lambda,-\sigma),$
the potential in (\ref{Lax.2}) remains invariant. Therefore the
eigenfunctions $\psi(x,\lambda,\sigma)$ and
$\psi(x,-\lambda,-\sigma)$ are solutions to the same spectral
problem. Since the asymptotics of these solutions do not depend on
$\sigma$, it follows that
\begin{eqnarray}\label{eigenbasis.real.lambda}
    \nonumber   \psi_{1}(x,\lambda,\sigma) &=& \psi_{2}(x,-\lambda,-\sigma)\\
                \phi_{1}(x,\lambda,\sigma)  &=& \phi_{2}(x,-\lambda,-\sigma),
\end{eqnarray}
Thus, we can write the two bases using just one of the functions,
say  $\psi(x,\lambda,\sigma)\equiv \psi_{1}(x,\lambda,\sigma)$ and
$\phi(x,\lambda,\sigma)\equiv \phi_{1}(x,\lambda,\sigma)$ as
$\psi_{2}(x,\lambda,\sigma)=\psi(x,-\lambda,-\sigma)$ and
$\phi_2(x,\lambda,\sigma)\equiv \phi(x,-\lambda,-\sigma)$.

When $u(x,t)$ and $\eta(x,t)$ are real, the spectral problem
(\ref{Lax.2}) is invariant under $\mathbb{Z}_2$ reduction group
\cite{M}, i.e. it has the following property: if
$\psi(x,\lambda,\sigma)$ is an eigenfunction, so is
$\bar{\psi}(x,\bar{\lambda},\sigma)$. Comparing the asymptotics
again, we conclude that this coinsides with the second Jost
solution, i.e.\begin{equation} \bar{\psi}(x,\bar{\lambda},\sigma)=
\psi(x,-\lambda,-\sigma) \label{Z2}
\end{equation}

Thus, for $\lambda \in \R$ we also have $\psi(x,\lambda,\sigma) =
\bar{\psi}(x,-\lambda,-\sigma)$. From this, and
(\ref{scattering.eqn}) it follows that the scattering matrix
$\mathbf{T}(\lambda)$ may be written in the form
\begin{equation}
    \mathbf{T}(\lambda,\sigma) = \left(
                            \begin{array}{cc}
                                    a(\lambda,\sigma) & b(\lambda,\sigma) \\
                                    \bar{b}(\lambda,\sigma) & \bar{a}(\lambda,\sigma)
                            \end{array}
                          \right),
\end{equation}
for spectral parameter $\lambda \in \mathbb{R}.$

We now have the following relationship between $\phi(x,\lambda,\sigma)$ and the Jost solutions $\psi(x,\lambda,\sigma), \bar{\psi}(x,\lambda,\sigma),$
\begin{equation}\label{phi.psi}
    \phi(x,\lambda,\sigma)  = a(\lambda,\sigma)\psi(x,\lambda,\sigma) + b(\lambda,\sigma)\psi(x,-\lambda,-\sigma).
\end{equation}
Furthermore, for any pair of solutions $f_{1}$ and $f_{2}$ to (\ref{Lax.2}) the Wronskian of the pair is independent of $x$,
\[\partial_{x}W[f_{1},f_{2}] = \partial_{x}(f_{1}\partial_{x}f_{2} - f_{2}\partial_{x}f_{1}) = 0.\]
In particular, it follows that the Jost solutions satisfy the following condition
\begin{equation}\label{Wronskian}
    W[\phi(x,\lambda,\sigma),\bar{\phi}(x,\lambda,\sigma)] = W[\psi(x,\lambda,\sigma),\bar{\psi}(x,\lambda,\sigma)] = 2i\lambda,
\end{equation}
which clearly follows from the asymptotic behaviour of $\{\psi(x,\lambda,\sigma)\bar{\psi}(x,\lambda,\sigma)\}$ and $\{\phi(x,\lambda,\sigma),\bar{\phi}(x,\lambda,\sigma)\}$ as $|x|\to \infty.$
It follows from (\ref{phi.psi}) and (\ref{Wronskian}) that
\begin{equation}\label{det}
    \det\mathbf{T(\lambda,\sigma)} = |a(\lambda,\sigma)|^{2} - |b(\lambda,\sigma)|^{2} = 1, \qquad \lambda \in \mathbb{R}.
\end{equation}

\section{Asymptotic behaviour of the Jost solutions}

Since the functions $u(x)$ and $w(x)$ are Schwartz class it follows that the solution $\psi(x,\lambda,\sigma)$ have asymptotic behaviour such that
\begin{equation}
    \psi_{xx}(x,\lambda,\sigma) \to -\lambda^{2}e^{-i\lambda x}, \quad x \to +\infty.
\end{equation}
Consequently, we make the following ansatz for the asymptotic
expansion as $|\lambda| \to \infty$,
\begin{equation}\label{ansatz}
    \psi(x,\lambda,\sigma) =
\left[X_{0}(x,\sigma) + \frac{1}{\lambda}X_{1}(x,\sigma) +
\mathcal{O}(\lambda^{-2})\right]e^{-i\lambda x},
\end{equation}
where the function $X_{0}(x,\sigma)$ and $X_{1}(x,\sigma)$ behave
asymptotically according to
\begin{equation}\label{asymptotics.xs}
    \begin{array}{c}
      X_{0}(x,\sigma) \to 1 \\
      X_{1}(x,\sigma) \to 0
    \end{array} \Bigg\}, \quad x \to +\infty.
\end{equation}
The substitution of (\ref{ansatz}) into (\ref{Lax.2}) gives
\begin{eqnarray}
\frac{\partial_{x}X_{0}(x,\sigma)}{X_{0}(x,\sigma)} &=& \frac{i}{2}\sigma u(x),\label{X0}\\
\sigma u(x) X_{1}(x,\sigma) + 2i \partial_{x}X_{1}(x,\sigma) &=&
-w(x)\cdot X_{0}(x,\sigma) + \partial_{x}^{2}X_{0}(x,\sigma).
\label{X1}
\end{eqnarray}
Using the conditions in (\ref{asymptotics.xs}) we may easily solve
(\ref{X0}), (\ref{X1}) to give the following expressions for
$X_{0}(x,\sigma)$ and $X_{1}(x,\sigma)$:
\begin{eqnarray}\label{x0.x1}
    \nonumber   X_{0}(x,\sigma) &=& \exp\left\{-\frac{i\sigma}{2}\int_{x}^{\infty}u(x')dx'\right\},\\
                X_{1}(x,\sigma) &=& X_{0}(x,\sigma)\cdot\left[\frac{\sigma}{4}u(x) - \frac{i}{8}\int_{x}^{\infty}(u^{2}(x') + 4w(\xi))dx'\right].
\end{eqnarray}
Similarly, we obtain analogous expressions for
$\psi(x,\lambda,\sigma)$, i.e.
\begin{eqnarray}
\psi(x,\lambda,\sigma) &=& e^{-i\left(\lambda x +
\frac{\sigma}{2}\int_{x}^{\infty}u(x')dx'\right)}\left[1
+ \frac{1}{\lambda}\xi_{1}(x,\sigma)) +\ldots\right],\label{expansion.basis1}\\
\phi(x,\lambda,\sigma) &=&    e^{-i\left(\lambda x -
\frac{\sigma}{2}\int^{x}_{-\infty}u(x')dx'\right)}\left[1 +
\frac{1}{\lambda}\zeta_{1}(x,\sigma) +\ldots\right],
\label{expansion.basis2}
\end{eqnarray}
where the functions $\xi_{1}(x)$ and $\zeta_{1}(x)$ are given by
\begin{eqnarray}\label{xi.zeta}
      \nonumber\xi_{1}(x,\sigma) &=& \frac{\sigma}{4}u(x) - \frac{i}{8}\int^{\infty}_{x}(u^{2}(x') + 4w(x'))dx',\\
      \zeta_{1}(x,\sigma) &=& \frac{\sigma}{4}u(x) + \frac{i}{8}\int_{-\infty}^{x}(u^{2}(x') + 4w(x'))dx'.
\end{eqnarray}

\section{Analytic behaviour of the Jost solutions}
We now define the related function
\begin{equation}\label{chi.1}
      \chi^{(+)}(x,\lambda,\sigma) = e^{i\lambda x}\phi(x,\lambda,\sigma) \to 1, \quad x \to -\infty.
\end{equation}
Using
\[e^{i\lambda x}\phi_{x}(x,\lambda,\sigma) = \chi_{x}^{(+)}(x,\lambda,\sigma) - i\lambda\chi^{(+)}(x,\lambda,\sigma)\]
along with the spectral problem in (\ref{Lax.2}),
we may write
\begin{equation}\label{chi1.xx}
    \chi_{xx}^{(+)}(x,\lambda,\sigma) = (\lambda\sigma u(x) + w(x))\chi^{(+)}(x,\lambda,\sigma) + 2i\lambda\chi_{x}^{(+)}(x,\lambda,\sigma).
\end{equation}
Meanwhile the asymptotic expansion in $\lambda$ appearing in
(\ref{expansion.basis2}) suggests the following integral
representation for $\chi^{(+)}(x,\lambda,\sigma),$
\begin{equation}\label{chi.2}
\chi^{(+)}(x,\lambda,\sigma) = 1 +
\int_{-\infty}^{x}\frac{e^{2i\lambda(x-x')} -
1}{2i\lambda}P(x',\lambda,\sigma)\chi^{(+)}(x',\lambda,\sigma)dx'
\end{equation}
for some $P(x,\lambda,\sigma)\in \mathcal{S}(\mathbb{R})$. From
this integral representation differentiating twice we obtain
\begin{eqnarray}\label{chi2.xx}
\nonumber\chi^{(+)}_{xx}(x,\lambda,\sigma) &=&
P(x,\lambda,\sigma)\chi^{(+)}(x,\lambda,\sigma) +
2i\lambda\chi^{(+)}_{x}(x,\lambda,\sigma).
\end{eqnarray}
From (\ref{chi1.xx}) and (\ref{chi2.xx}) we determine
\begin{equation}\label{pxl}
P(x,\lambda,\sigma) = \lambda\sigma u(x) + w(x),
\end{equation}
and so we may write
\begin{equation}\label{chi.3}
\chi^{(+)}(x,\lambda,\sigma) = 1 +
\int_{-\infty}^{x}\frac{e^{2i\lambda(x-x')} -
1}{2i\lambda}(\lambda\sigma u(x') +
w(x'))\chi^{(+)}(x',\lambda,\sigma)dx'.
\end{equation}
Of particular importance and clear from (\ref{chi.3}) is the
analytic properties of $\chi^{(+)}(x,\lambda,\sigma)$. We can see
that for all values of $x$ the kernel of the integral above is
finite for all values of $\lambda$ such that $\mbox{Im }\lambda >
0.$ Therefore $\chi^{(+)}(x,\lambda,\sigma)$ and
$\phi(x,\lambda,\sigma)$ are  analytic in the upper half plane
$\mathbb{C}_{+}$. It obviously follows that
$\bar{\chi}^{(+)}(x,\bar{\lambda},\sigma)$ is analytic for
$\lambda \in \mathbb{C}_{-}$.

In a similar manner we may define
\begin{equation}\label{chim.1}
\chi^{(-)}(x,\lambda,\sigma) = e^{i\lambda
x}\psi(x,\lambda,\sigma)  \to 1, \quad x \to +\infty
\end{equation}
from which it follows that
\begin{equation}\label{chim.2}
\chi^{(-)}(x,\lambda,\sigma) = 1 -
\int_{x}^{\infty}\frac{e^{2i\lambda(x-x')} -
1}{2i\lambda}(\lambda\sigma u(x) +
w(x))\chi^{(-)}(x,\lambda,\sigma).
\end{equation}
It is immediately clear from (\ref{chim.2}) that
$\chi^{(-)}(x,\lambda,\sigma)$ and therefore
$\psi(x,\lambda,\sigma)$ are analytic throughout $\mathbb{C}_{-}.$

Next we introduce new notation for later convenience,
\begin{equation}\label{gamma}
\omega_{-}(x) = \frac{1}{2}\int_{-\infty}^{x}u(x')dx' \quad
\mbox{and} \quad \omega_{+}(x) =
\frac{1}{2}\int_{x}^{\infty}u(x')dx'.
\end{equation}
With this we may rewrite (\ref{expansion.basis1}),
(\ref{expansion.basis2}) as follows
\begin{eqnarray}\label{underlined}
\nonumber\underline{\psi}(x,\lambda,\sigma) =
\psi(x,\lambda,\sigma)e^{i(\lambda x + \sigma\omega_{+}(x))}
= 1 + \frac{1}{\lambda}\xi_{1}(x,\sigma),\\
\underline{\phi}(x,\lambda,\sigma) =
\phi(x,\lambda,\sigma)e^{i(\lambda x - \sigma\omega_{-}(x))} = 1 +
\frac{1}{\lambda}\zeta_{1}(x,\sigma).
\end{eqnarray}
To obtain the analytic properties of
$\underline{\psi}(x,\lambda,\sigma)$ and
$\underline{\phi}(x,\lambda,\sigma),$ we note that
\[\underline{\psi}(x,\lambda,\sigma) =
\chi^{(-)}(x,\lambda,\sigma)e^{i\sigma\omega_{+}(x)}
\quad\mbox{and}\quad \underline{\phi}(x,\lambda,\sigma) =
\chi^{(+)}(x,\lambda,\sigma)e^{-i\sigma\omega_{-}(x)}. \] Since
$u(x)$ is Schwartz class and independent of $\lambda$ and given
the analyticity of $\chi^{(\pm)}(x,\lambda,\sigma)$ throughout
$\mathbb{C}_{\pm}$ respectively, it follows that
$\underline{\phi}(x,\lambda,\sigma)$ and
$\underline{\psi}(x,\lambda,\sigma)$ are also analytic throughout
$\mathbb{C}_{+}$ and $\mathbb{C}_{-}$ respectively.

\section{The $t$-dependence of the scattering data}
We may rewrite the second member of the Lax pair in terms of the
auxiliary function $u(x)$ and add an arbitrary constant $\gamma,$
without effecting the physical equations of motion, to obtain,
\begin{equation}\label{td_1}
\Psi_{t}(x,\lambda,\sigma) = -\left(\sigma\lambda +
\frac{1}{2}u(x)\right)\Psi_{x}(x,\lambda,\sigma) +\left(\gamma +
\frac{1}{4}u_{x}(x)\right)\Psi(x,\lambda,\sigma).
\end{equation}
In particular we may write
\begin{equation}\label{dt_phi}
\phi_{t}(x,\lambda,\sigma) = -\left(\sigma\lambda +
\frac{1}{2}u(x)\right)\phi_{x}(x,\lambda,\sigma) +\left(\gamma +
\frac{1}{4}u_{x}(x)\right)\phi(x,\lambda,\sigma).
\end{equation}
However, we also note that along the discrete spectrum we have the
scattering relation (\ref{phi.psi}), from which we may obtain the
asymptotic behavior of $\phi_{t}(x,\lambda,\sigma)$ as $x \to
+\infty,$ namely
\begin{equation} \label{phit1}
\phi_{t}(x,\lambda,\sigma) \to a_{t}(\lambda,\sigma)e^{-i\lambda
x} + b_{t}(\lambda,\sigma)e^{+i\lambda x}.
\end{equation}
Using the r.h.s of (\ref{dt_phi}) along with (\ref{phi.psi}), we find as $x \to +\infty$ that
\begin{eqnarray}
\phi_{t}(x,\lambda,\sigma)\to& -& \sigma\lambda[-i\lambda
a(\lambda,\sigma)e^{-i\lambda x}+ i\lambda
b(\lambda,\sigma)e^{i\lambda x}]  \nonumber \\
&+ & \gamma[a(\lambda,\sigma)e^{-i\lambda x} +
b(\lambda,\sigma)e^{+i\lambda x}], \label{phit2}
\end{eqnarray}
where we have made use of the fact that $u(x)$ is Schwartz class
and vanishes when $x\to \pm \infty$. Making the choice $\gamma =
-i\sigma\lambda^{2},$ the $t$-derivative of $a(\lambda, \sigma)$
vanishes.  It follows that we may write
\begin{equation}
\begin{array}{c}
a_{t}(\lambda,\sigma) = 0 \Rightarrow a(\lambda,\sigma,t) = a(\lambda,\sigma,0),\\
b_{t}(\lambda,\sigma) = -2i\lambda^{2}b(\lambda,\sigma)
\Rightarrow b(\lambda,\sigma,t) =
b(\lambda,\sigma,0)e^{-2i\sigma\lambda^{2}t}. \label{b}
\end{array}
\end{equation}

Along the discrete spectrum, we have $a(\lambda_{n},\sigma) = 0$
and therefore instead of (\ref{phit1}) we have
\begin{equation}
\phi_{t}(x,\lambda_n,\sigma) \to  b_{n,t}(\sigma)e^{i\lambda_n x}.
\end{equation} Instead of (\ref{phit2}) we have
\begin{equation}
\phi_{t}(x,\lambda_n,\sigma) \to -\sigma\lambda_n[i\lambda_n
b_n(\sigma)e^{i\lambda_n x}] + \gamma(\lambda_n)
b_n(\sigma)e^{i\lambda_n x},
\end{equation}
and thus
\begin{equation}
b_{n,t}(\sigma) =- 2i\sigma\lambda_{n}^{2}b_{n}(\sigma),
\end{equation}
from where
\begin{equation}
b_{n}(\sigma,t) =
b_{n}(\sigma,0)e^{-2i\sigma\lambda_{n}^{2}t}.\label{bn}
\end{equation}

\section{Conservation Laws}
We may derive a collection of conserved quantities from the
spectral problem introduced in the previous section in
(\ref{Lax.2}). To proceed we first introduce the function
\begin{equation}\label{rho.def}
    \rho(x,\lambda,\sigma) = \frac{\Psi_{x}(x,\lambda,\sigma)}{\Psi(x,\lambda,\sigma)}.
\end{equation}
Differentiating once with respect to $x$ we find
\begin{equation} \label{rho.eq}
    \rho^{2} + \rho_{x} = -\lambda^{2} + \sigma\lambda u  + w.
\end{equation}
Using this result, along with the Lax pair in (\ref{Lax.2}) and
(\ref{td_1}) we find upon differentiating (\ref{rho.def}) with
respect to $t$ that
\begin{equation}{\label{rho.t}}
\rho_{t} = \frac{1}{4}\left(u_{x} - 2\rho u - 4 \sigma\lambda \rho
\right)_{x}.
\end{equation}
Using the the fact $u(x)$ and $w(x)$ are Schwartz class, we see
from (\ref{rho.t}) that
\begin{equation}
    \displaystyle{\int_{-\infty}^{+\infty}}\rho_{t}(x,\lambda,\sigma)dx = 0,
\end{equation}
that is to say
\begin{equation}\label{cq.1}
    \mathcal{I}_{0} (\lambda) = \displaystyle{\int_{-\infty}^{+\infty}}\rho(x,\lambda,\sigma)dx
\end{equation}
is a generating function for the conserved quantities. We may
expand it in a power series in $\lambda$ according to
\begin{equation}\label{expansion.0}
    \mathcal{I}_{0} = \lambda\mathcal{I}_{1} + \mathcal{I}_{2} +
    \frac{\mathcal{I}_{3}}{\lambda} +
    \frac{\mathcal{I}_{4}}{\lambda^{2}} + \mathcal{O}(\lambda^{-3}),
\end{equation}
where $\mathcal{I}_{1}, \mathcal{I}_{2},$ etc is an infinite
sequence of conserved quantities. Next, we expand
$\rho(x,t,\lambda)$ as a power series in $\lambda$
\begin{equation}\label{expansion.1}
    \rho(x,\lambda,\sigma) = i\lambda + \rho_{0}(x) + \frac{\rho_{1}(x,t)}{\lambda} + \frac{\rho_{2}(x)}{\lambda^{2}} + \mathcal{O}(\lambda^{-3}),
\end{equation}
and use it in (\ref{rho.eq}), then the terms of equivalent order
in $\lambda$ give
\begin{equation}\label{cg.0}
    \rho_{0}(x) = -\frac{i\sigma}{2}u(x)
\end{equation}
and as a result of (\ref{expansion.0}) it follows that
\begin{equation}\label{int.0}
    \mathcal{I}_{2} = \int_{-\infty}^{+\infty}\rho_0(x)dx = -\frac{i\sigma}{2}\int_{-\infty}^{+\infty}u(x)dx.
\end{equation}
So we see that
\begin{equation}
   \alpha_1 \equiv \frac{1}{2} \int_{-\infty}^{+\infty}u(x)dx
   \label{alpha1}
\end{equation}
is an integral of motion. Following a similar procedure, we find
the next conserved quantities to be
\begin{eqnarray}
    \nonumber\mathcal{I}_{3} = -\frac{i}{8}\int_{-\infty}^{+\infty}(u^{2}(x) + 4w(x))dx,\\
    \mathcal{I}_{4} = -\frac{i\sigma}{16}\int_{-\infty}^{+\infty}u(x)(u^{2}(x) + 4w(x))dx.
\end{eqnarray}
One may continue a process of iteration indefinitely, whereby an
infinite series of such conserved quantities is generated from the
$u(x)$ and $w(x),$ and therefore from the physical variables
$u(x)$ and $\eta(x)$.

\section{Analytic continuation of $a(\lambda, \sigma)$}
Returning to (\ref{phi.psi}) we see that we may re-write the
scattering coefficient $a(\lambda,\sigma)$ in terms of the
$x$-independent Wronsikian,
\begin{equation}\label{a-lambda}
a(\lambda,\sigma) =
\frac{W[\phi(x,\lambda,\sigma),\psi(x,-\lambda,-\sigma)]}{2i\lambda}.
\end{equation}
Since the two eigenfunctions  in (\ref{a-lambda}) are analytic for
$\lambda \in \mathbb{C}_+$, $a(\lambda, \sigma)$ allows an
analytic continuation in the upper half complex plane. From
(\ref{a-lambda}) with (\ref{expansion.basis1})--
(\ref{expansion.basis2}) we obtain the asymptotic behavior of the
scattering coefficient,
\begin{equation}\label{a-asymptotic}
    \lim_{|\lambda|\to\infty}a(\lambda,\sigma) = e^{i\sigma\alpha_{1}}
\end{equation} where $\alpha_1$ is the conserved quantity
(\ref{alpha1}). We make further the assumption that
$a(\lambda,\sigma)$ has a finite number of \textit{simple} zeros
$\lambda_{n} \in  \C_{+}$,  $n = 1,2,3,\ldots,N$. We introduce the
auxiliary function
\begin{equation}\label{analytic_a.1}
A(\lambda,\sigma) =
e^{-i\sigma\alpha_{1}}\displaystyle{\prod_{n=1}^{N}}\frac{\lambda
- \bar{\lambda}_{n}}{\lambda - \lambda_{n}}a(\lambda,\sigma),
\end{equation}
which is analytic without zeroes in  $\C_{+}$. It follows from
(\ref{analytic_a.1}) that
\begin{equation}\label{A_a}
|A(\lambda,\sigma)| = |a(\lambda,\sigma)|, \quad \lambda \in \R,
\end{equation}
Next, we also see from (\ref{a-asymptotic}) and (\ref{analytic_a.1}) that
\begin{equation}\label{lim_A}
\displaystyle{\lim_{|\lambda|\to\infty}}\ln A(\lambda,\sigma) = 0,
\end{equation}
and so, $\ln A(\lambda,\sigma)$ is analytic throughout $\C_{+}$
and vanishes as $|\lambda| \to \infty.$

We also have from (\ref{A_a})
\[\ln A(\lambda,\sigma) = \ln|A(\lambda,\sigma)| + i\arg A(\lambda,\sigma) =
\ln|a(\lambda,\sigma)| + i\arg A(\lambda,\sigma),\] for $\lambda
\in \R.$ We make use of the Kramers-Kronig dispersion relations,
\begin{eqnarray}\label{Kramers-Kronig}
\nonumber   \ln|a(\lambda,\sigma)| = \frac{1}{\pi}\fint\limits_{-\infty}^{\infty}\frac{\arg A(\lambda',\sigma)}{\lambda' - \lambda}d\lambda'\\
\arg{A(\lambda,\sigma)} =
-\frac{1}{\pi}\fint\limits_{-\infty}^{\infty}\frac{\ln|a(\lambda',\sigma)|}{\lambda'
- \lambda}d\lambda',
\end{eqnarray}
for $\lambda \in \R,$ where the dashed integral denoted the
principal value part of the integral. Then with
(\ref{Kramers-Kronig}) we have
\begin{eqnarray} \ln A(\lambda,\sigma) &= &\ln|a(\lambda,\sigma)|
-\frac{i}{\pi}\fint\limits_{-\infty}^{\infty}\frac{\ln|a(\lambda',\sigma)|}{\lambda'
- \lambda}d\lambda' \nonumber \\
& = & \frac{1}{\pi
i}\int\limits_{-\infty}^{\infty}\frac{\ln|a(\lambda',\sigma)|}{\lambda'
- \lambda - i0^{+}}d\lambda', \qquad \lambda \in \mathbb{R}.
\label{lnA1}
\end{eqnarray}
Meanwhile, (\ref{analytic_a.1}) gives
\begin{equation}\label{lnA2}
    \ln A(\lambda,\sigma) = -i\sigma\alpha_{1} - \displaystyle\sum_{n=1}^{N}\frac{\lambda - \lambda_{n}}{\lambda - \bar{\lambda}_{n}} + \ln a(\lambda,\sigma).
\end{equation}
Using (\ref{lnA1}) and (\ref{lnA2}), we find that for real values
of $\lambda$ we may write
\begin{equation}\label{lna}
\ln a(\lambda,\sigma) = i\sigma\alpha_{1} +
\displaystyle\sum_{n=1}^{N}\frac{\lambda - \lambda_{n}}{\lambda -
\bar{\lambda}_{n}} + \frac{1}{\pi
i}\int\limits_{-\infty}^{\infty}\frac{\ln|a(\lambda',\sigma)|}{\lambda'
- \lambda - i0^{+}}d\lambda'
\end{equation} and for $\lambda \in \mathbb{C}_+$ the analytical
continuation is
\begin{equation}\label{lnac+}
\ln a(\lambda,\sigma) = i\sigma\alpha_{1} +
\displaystyle\sum_{n=1}^{N}\frac{\lambda - \lambda_{n}}{\lambda -
\bar{\lambda}_{n}} + \frac{1}{\pi
i}\int\limits_{-\infty}^{\infty}\frac{\ln|a(\lambda',\sigma)|}{\lambda'
- \lambda }d\lambda'.
\end{equation}

\section{The Riemann-Hilbert problem}

We may re-write the expression (\ref{phi.psi}) in terms of the new
analytic functions
$\underline{\phi}(x,\lambda,\sigma),\underline{\psi}(x,\lambda,\sigma)$
using (\ref{Z2}) for $\lambda \in \mathbb{R},$ as follows
\begin{equation}\label{underline-phi-psi}
\frac{\underline{\phi}(x,\lambda,\sigma)e^{i\sigma\alpha_{1}}}{a(\lambda,\sigma)}=
\underline{\psi}(x,\lambda\,\sigma) +
r(\lambda,\sigma)\underline{\bar{\psi}}(x,\lambda,\sigma)e^{2i(\lambda
x + \sigma\omega_{+}(x))},
\end{equation}  $r(\lambda,\sigma)=b(\lambda,\sigma)/a(\lambda,\sigma)$. The function
$\frac{\underline{\phi}(x,\lambda,\sigma)e^{i\sigma\alpha_{1}}}{a(\lambda,\sigma)}$
is analytic for $\mathrm{Im}\phantom{*} \lambda>0$, while
$\underline{\psi}(x,\lambda\,\sigma)$ is analytic for
$\mathrm{Im}\phantom{*} \lambda <0$. Thus, equation
(\ref{underline-phi-psi}) represents an additive Riemann-Hilbert
Problem (RHP) with a jump on the real line, given by
$$r(\lambda,\sigma)\underline{\bar{\psi}}(x,\lambda,\sigma)e^{2i(\lambda
x + \sigma\omega_{+}(x))}$$ and a normalization condition
$\lim_{|\lambda|\to
\infty}\underline{\psi}(x,\lambda,\sigma)=X_0(x,\sigma)$.

In this section we will follow the standard technique for solving
RHP. We integrate the two analytic functions with respect to
$\oint \frac{d \lambda'}{\lambda'-\lambda} (\cdot)$ over the
boundary of their analyticity domains, using the normalization
condition. In our case the domains (the upper $\mathbb{C}_{+}$ and
the lower $\mathbb{C}_{-}$ complex half-planes) have the real line
as a common boundary and there we relate the integrals using the
jump condition. The RHP approach for various equation is presented
in \cite{gvy2008,Lap07,HI2011,SS97}.

We now choose some $\lambda \in \mathbb{C}_{-}$ and integrate the left-hand side as follows,
\begin{equation}
\frac{1}{2\pi
i}\ointctrclockwise_{C^{+}}\frac{\underline{\phi}(x,\lambda',\sigma)e^{i\sigma\alpha_{1}}}{a(\lambda',\sigma)
\cdot(\lambda'-\lambda)}d\lambda'
=\sum_{n=1}^{N}\frac{\underline{\phi}^{(n)}(x,\sigma)e^{i\sigma\alpha_{1}}}{\dot{a}_{n}(\sigma)\cdot(\lambda_{n}
- \lambda)}
\end{equation}
where $C^{+}$ is the contour in the upper half plane shown in Fig.
1, $$\dot{a}_{n}(\sigma)\equiv \left(\frac{\text{d}a(\lambda,
\sigma)}{\text{d}\lambda}\right)_{\lambda=\lambda_n}\neq 0,\qquad
\underline{\phi}^{(n)}(x,\sigma)\equiv
\underline{\phi}(x,\lambda_n,\sigma).$$
\begin{figure}
\begin{center}
\includegraphics[width = 8cm]{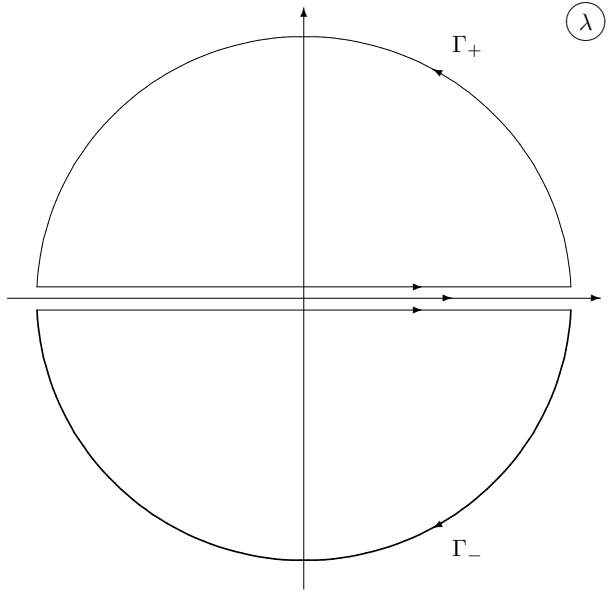}
\end{center}
\caption{The integration contours $C^{+}$ and $C^{-}$ are the
closed paths in the upper and lower half planes correspondingly;
$\Gamma_{\pm}$ are the semicircles with an infinite radius.}
\end{figure} We may write the integral as such because $\lambda\in \mathbb{C}_{-}$ and so
$\frac{1}{\lambda - \lambda'}$ is analytic throughout
$\mathbb{C}_{+}$. Furthermore, $a(\lambda)$ is analytic with
finite number of simple zeros, $\lambda_{n}$ in $\mathbb{C}_{+},$
and the function $\underline{\phi}(x,\lambda)$ is analytic
throughout $\mathbb{C}_{+}.$ Alternatively we may expand the
integral as follows
\begin{eqnarray}
    \nonumber\frac{1}{2\pi i}\ointctrclockwise_{C^{+}}\frac{\underline{\phi}(x,\lambda',\sigma)e^{i\alpha_{1}}}{a(\lambda',\sigma)\cdot(\lambda'-\lambda)}d\lambda' &=&\\ \frac{1}{2\pi i}\int_{-\infty}^{\infty}\frac{\underline{\phi}(x,\lambda',\sigma)e^{i\alpha_{1}}}{a(\lambda')\cdot(\lambda'-\lambda)}d\lambda' &+&
    \frac{1}{2\pi i}\int_{\Gamma_{+}}\frac{\underline{\phi}(x,\lambda',\sigma)e^{i\alpha_{1}}}{a(\lambda',\sigma)\cdot(\lambda'-\lambda)}d\lambda'.
\end{eqnarray}
Using the asymptotic properties of $a(\lambda,\sigma)$ and
$\underline{\phi}(x,\lambda,\sigma)$ along with the relationship
(\ref{underline-phi-psi}), we find
\begin{eqnarray}\label{RHP1}
\sum_{n=1}^{N}\frac{\underline{\phi}^{(n)}(x,\sigma)e^{i\alpha_{1}}}{\dot{a}(\lambda_{n})\cdot(\lambda_{n}
- \lambda)} &=& \frac{1}{2\pi
i}\int_{\Gamma^{+}}\frac{1}{\lambda'-\lambda}d\lambda' +
\frac{1}{2\pi
i}\int_{-\infty}^{\infty}\frac{\underline{\psi}(x,\lambda',\sigma)}{\lambda'
- \lambda}d\lambda'   \nonumber \\
&+& \frac{1}{2\pi
i}\int_{-\infty}^{\infty}\frac{r(\lambda',\sigma)e^{2i(\lambda'x+\sigma\omega_{+}(x))}
\underline{\bar{\psi}}(x,\lambda',\sigma)}{\lambda'
- \lambda}d\lambda'.\nonumber \\
\end{eqnarray}

Next we obtain an expression for the line-integral
\[\frac{1}{2\pi i}\int_{-\infty}^{+\infty}\frac{\underline{\psi}(x,\lambda',\sigma)}{\lambda'-\lambda}d\lambda',\]
by considering the integral over the contour $C^{-},$ shown in
Fig. 1. Since $\lambda \in \mathbb{C}_{-}$ and
$\underline{\psi}(x,\lambda,\sigma)$ is analytic therein. In
addition the contour $C^{-}$ is clockwise, so it follows that
\begin{equation}
\frac{1}{2\pi
i}\ointclockwise_{C^{-}}\frac{\underline{\psi}(x,\lambda,\sigma)}{\lambda'-\lambda}d\lambda'
= -\psi(x,\lambda,\sigma).
\end{equation}
Expanding the integral, we have
\begin{equation}
    \frac{1}{2\pi i}\ointclockwise_{C^{-}}\frac{\underline{\psi}(x,\lambda',\sigma)}{\lambda'-\lambda}d\lambda' = \frac{1}{2\pi i}\int_{-\infty}^{+\infty}\frac{\underline{\psi}(x,\lambda',\sigma)}{\lambda'-\lambda}d\lambda' + \frac{1}{2\pi i}\int_{\Gamma_{-}}\frac{\underline{\psi}(x,\lambda',\sigma)}{\lambda'-\lambda}d\lambda'.
\end{equation}
Using the asymptotic properties of
$\underline{\psi}(x,\lambda,\sigma)$ as $|\lambda| \to \infty$
with $\lambda \in \mathbb{C}^{-},$ we have,
\begin{equation}\label{RHP2}
\frac{1}{2\pi
i}\int_{-\infty}^{\infty}\frac{\underline{\psi}(x,\lambda',\sigma)}{\lambda'-\lambda}d\lambda'
= -\underline{\psi}(x,\lambda,\sigma) - \frac{1}{2\pi
i}\int_{\Gamma_{-}}\frac{1}{\lambda' - \lambda}d\lambda'
\end{equation}
We can also make use of the following result when it comes to
substituting this expression in (\ref{RHP1}),
\begin{equation}
\int_{\Gamma_{+}}\frac{1}{\lambda' - \lambda}d\lambda' -
\int_{\Gamma_{-}}\frac{1}{\lambda' - \lambda}d\lambda' = 2\pi i.
\end{equation}
Upon making these substitutions we find the following integral
representation for $\underline{\psi}(x,\lambda,\sigma)$, $\lambda
\in \mathbb{C}_-$:
\begin{equation}\label{RHP2}
\underline{\psi}(x,\lambda,\sigma) = 1 -
\sum_{n=1}^{N}\frac{\underline{\phi}^{(n)}(x,\sigma)e^{i\alpha_{1}}}{\dot{a}_{n}(\sigma)(\lambda_{n}
- \lambda)} + \frac{1}{2\pi
i}\int_{-\infty}^{\infty}\frac{r(\lambda')e^{2i(\lambda'x +
\sigma\omega_{+}(x))}\underline{\bar{\psi}}(x,\lambda',\sigma)}{\lambda'-\lambda}d\lambda'.
\end{equation}
Since at the points of the discrete spectrum
$\phi(x,\lambda_n,\sigma)=b_n(\sigma)\bar{ \psi}(x,
\bar{\lambda}_n, \sigma)$ we have
\begin{equation}
\frac{\underline{\phi}^{(n)}(x,\sigma)e^{i\alpha_{1}}}{\dot{a}_{n}(\sigma)}
= iR_{n}(\sigma)e^{2i(\lambda_{n}x +
\sigma\omega_{+}(x))}\underline{\bar{\psi}}(x,\bar{\lambda}_n,\sigma)
\end{equation}
where we define
\[R_{n}(\sigma) = \frac{b_{n}(\sigma)}{i\dot{a}_{n}(\sigma)}.\]
The Riemann-Hilbert problem is reduced to the linear singular
integral equation for $\underline{\psi}(x,\lambda,\sigma)$
\begin{eqnarray}\label{Riemann-Hilbert}
\nonumber\underline{\psi}(x,\lambda,\sigma) = 1 -
i\sum_{n=1}^{N}\frac{R_{n}(\sigma)\underline{\bar{\psi}}(x,\bar{\lambda}_n,\sigma)}{(\lambda_{n}
- \lambda)}e^{2i(\lambda_{n}x + \sigma \omega_{+}(x))} \\+
\frac{1}{2\pi
i}\int_{-\infty}^{\infty}\frac{r(\lambda',\sigma)\underline{\bar{\psi}}(x,\lambda',\sigma)}{\lambda'-\lambda}e^{2i(\lambda'x
+ \sigma\omega_{+}(x))}d\lambda'.
\end{eqnarray}

In addition to (\ref{Riemann-Hilbert})  we have an analogous
system written at the points $\lambda=\bar{\lambda}_p \in
\mathbb{C}_-$, $p=1,2, \ldots, N$:

\begin{eqnarray}\label{Riemann-Hilbert1}
\nonumber\underline{\psi}(x,\bar{\lambda}_p,\sigma) = 1 -
i\sum_{n=1}^{N}\frac{R_{n}(\sigma)\underline{\bar{\psi}}(x,\bar{\lambda}_n,\sigma)}{(\lambda_{n}
- \bar{\lambda}_p)}e^{2i(\lambda_{n}x + \sigma \omega_{+}(x))} \\+
\frac{1}{2\pi
i}\int_{-\infty}^{\infty}\frac{r(\lambda',\sigma)\underline{\bar{\psi}}(x,\lambda',\sigma)}
{\lambda'-\bar{\lambda}_p}e^{2i(\lambda'x +
\sigma\omega_{+}(x))}d\lambda'.
\end{eqnarray}

Finally, the fact that at $\lambda=0$ the Jost solution
$\psi(x,0,\sigma)$ does not depend on $\sigma $ gives
$\psi(x,0,\sigma)=\psi(x,0,-\sigma)$ or an algebraic system for
$e^{2i\sigma\omega_+(x)}$:

\begin{equation} \label{RHP3}
e^{2i\sigma\omega_+(x)}=\frac{\underline{\psi}(x,0,\sigma)}{\underline{\psi}(x,0,-\sigma)}
=\frac{\underline{\psi}(x,0,\sigma)}{\underline{\bar{\psi}}(x,0,\sigma)}.
\end{equation}

The system (\ref{Riemann-Hilbert}), (\ref{Riemann-Hilbert1}),
(\ref{RHP3}) allows for the determination of both the Jost
solution and the potential functions of the spectral problem in
terms of the scattering data. Note that the time-dependence of the
scattering data is known from (\ref{b}), (\ref{bn}):
\begin{equation} \label{SD4RHP} r(\lambda,\sigma,t) =
r(\lambda,\sigma,0)e^{-2i\sigma \lambda^2 t}, \qquad
R_n(\sigma,t)= R_n(\sigma,0)e^{-2i\sigma \lambda_n^2
t}.\end{equation}

Thus, the complete set of scattering data is

\begin{equation} \label{SD}
r(\lambda,\sigma,0), \quad \lambda_n, \quad R_n(\sigma,0)\quad
(n=1,2, \ldots, N).
\end{equation}

Also, it is sufficient to know the scattering data for $\sigma=1$,
because of the $\mathbb{Z}_2$ involution, which holds on the
scattering data too:
\begin{equation}r(\lambda,-\sigma)=\bar{r}(-\lambda,\sigma), \qquad
R_n(-\sigma)=\bar{R}_n(\sigma).
\end{equation}

\section{Reflectionless potentials and soliton solutions}

The so-called reflectionless potentials are a subclass which
corresponds to a restricted set of scattering data:
$r(\lambda,\sigma) = 0$; $\lambda \in \R$. Then the system
(\ref{Riemann-Hilbert}), (\ref{Riemann-Hilbert1}), (\ref{RHP3}) is
algebraic, and the solutions of the PDE are called solitons.

The simplest case is the $N=1$-soliton solution, so we start first
with this case. From (\ref{Riemann-Hilbert}) we have

\begin{eqnarray}\label{scalar_1}
\underline{\psi}(x;\lambda,\sigma) = 1 -
i\frac{R_{1}(\sigma)\underline{\bar{\psi}}
(x,\bar{\lambda}_{1},\sigma)}{\lambda_{1} -
\lambda}e^{2i(\lambda_{1}x+ \sigma\omega_{+}(x))},\qquad \lambda
\in \C_{-}.
\end{eqnarray}

We notice that $\bar{\psi} (x,\bar{\lambda},\sigma)$ has an unique
pole at $\bar{\lambda}_{1}$ and $\psi (x,-\lambda,-\sigma)$ has an
unique pole at $-\lambda_1$. Due to (\ref{Z2}) these two poles
coincide, i.e. $\bar{\lambda}_{1}=-\lambda_1$ and therefore
$\lambda_1=i\nu$ is purely imaginary, $\nu>0$ is real.

Solving for $\underline{\psi}(x,\bar{\lambda}_{1},\sigma)$ we find
\begin{equation}\label{psi_lambda_1}
\underline{\bar{\psi}}(x,\bar{\lambda}_{1},\sigma) = \frac{1 -
i\frac{\bar{R}_{1}(\sigma,t)}{2\lambda_{1}}e^{2i(\bar{\lambda}_{1}x+\sigma\omega_{+}(x))}}{1
+\frac{|R_{1}(\sigma,t)|^2 e^{4i\lambda_1 x}}{4\lambda_1^2}}.
\end{equation}
Then (\ref{scalar_1}) takes the form
\begin{equation}\label{psi-1}
\underline{\psi}(x,\lambda, \sigma ) = 1 + \frac{2 i
\nu}{\lambda-i\nu}\cdot\frac{\frac{R_1(\sigma, 0)}{2\nu}e^{-2\nu x
+2 i \sigma \nu^2 t+2i
\sigma\omega_{+}(x)}-\frac{|R_1(\sigma,0)|^2}{4\nu^2}e^{-4\nu
x}}{1-\frac{|R_{1}(\sigma,0)|^2}{4\nu^2}e^{-4\nu x}}
\end{equation}

Furthermore, we can relate the real and imaginary parts of the
complex constant $\frac{R_1(\sigma, 0)}{2\nu}$ to two new
constants, say $x_0$ and $t_0$ as follows:
$$\frac{R_1(\sigma, 0)}{2\nu}=e^{4 \nu x_0 -2i\sigma \nu^2 t_0}.$$

Now $\underline{\psi}(x,\lambda, \sigma )$ in (\ref{psi-1})
depends only on $x-x_0$, $t-t_0$ and due to the translational
invariance of the problem, without loss of generality, we can
choose $x_0=0$ and $t_0=0$. This simplifies (\ref{psi-1}) to
\begin{equation}\label{psi-2}
\underline{\psi}(x,\lambda, \sigma ) = 1 + \frac{2 i
\nu}{\lambda-i\nu}\cdot\frac{e^{-2\nu x +2 i \sigma \nu^2 t+2i
\sigma\omega_{+}(x)}-e^{-4\nu x}}{1-e^{-4\nu x}}
\end{equation}

Then (\ref{RHP3}) gives
\begin{equation}\label{exp-omega}
e^{2 i \sigma \omega_+(x,t)}=\frac{1+2e^{-2\nu x-2 i \sigma \nu^2
t}+e^{-4\nu x}}{1+2e^{-2\nu x+2 i \sigma \nu^2 t}+e^{-4\nu x}}
\end{equation}

From $\omega_+(x,t)$ and (\ref{gamma}) we can recover $u(x,t)$:

\begin{equation}\label{u}
u(x,t)=\nu \frac{\sin ( 2 \nu^2 t) \sinh  (2\nu x)}{\cosh^4 (\nu
x) \cos^2 (\nu^2 t )+ \sinh^4 (\nu x )\sin^2 (\nu^2 t)}
\end{equation}

On the other hand, we also have (\ref{underlined}),
\begin{equation}
\underline{\psi}(x,\lambda,\sigma) = 1 +
\frac{1}{\lambda}\left[\frac{\sigma}{4}u(x) +
\frac{i}{8}\int_{x}^{\infty}(u^{2}(x') + 4w(x'))dx'\right]+
\mathcal{O}\left(\frac{1}{\lambda^2}\right), \nonumber
\end{equation}



\noindent which can be compared to (\ref{psi-2}):

\begin{equation}\label{psi-3}
\underline{\psi}(x,\lambda, \sigma ) = 1 + \frac{2 i
\nu}{\lambda}\cdot\frac{e^{-2\nu x +2 i \sigma \nu^2 t+2i
\sigma\omega_{+}(x,t)}-e^{-4\nu x}}{1-e^{-4\nu x}}+
\mathcal{O}\left(\frac{1}{\lambda^2}\right).
\end{equation} Since $\omega_+(x,t)$ and $u(x,t)$ are already
known, we find $w(x,t)$ and finally $ \eta(x,t)$. With (\ref{w})
we compute $$ u^2+4w = 2(\kappa+\frac{1}{2})u^2+4 \eta.$$  For the
Kaup-Boussinesq case $ \kappa=-\frac{1}{2}$ and  $$ u^2+4w = 4
\eta= -4 \partial_x^2 \ln \left[ (1+e^{-2\nu x})^4+(1-e^{-2\nu
x})^4 \tan^2 \nu^2 t \right], $$

\begin{eqnarray}\nonumber
\eta\!=\!-\! 2\nu^2 \frac{\cosh^6(\nu x) \cos ^4
(\nu^2t)\!+\!\frac{3}{4}\sin^2 ( 2 \nu^2 t) \sinh^2 (2\nu
x)\!-\!\sinh^6(\nu x) \sin^4 (\nu^2 t)}{[\cosh^4 (\nu x) \cos^2
(\nu^2 t )\!+\! \sinh^4 (\nu x )\sin^2
(\nu^2 t)]^2} .\nonumber \\
\label{eta}\end{eqnarray}

\begin{figure}
\begin{tabular}{ccc}
\includegraphics[width=0.3\textwidth]{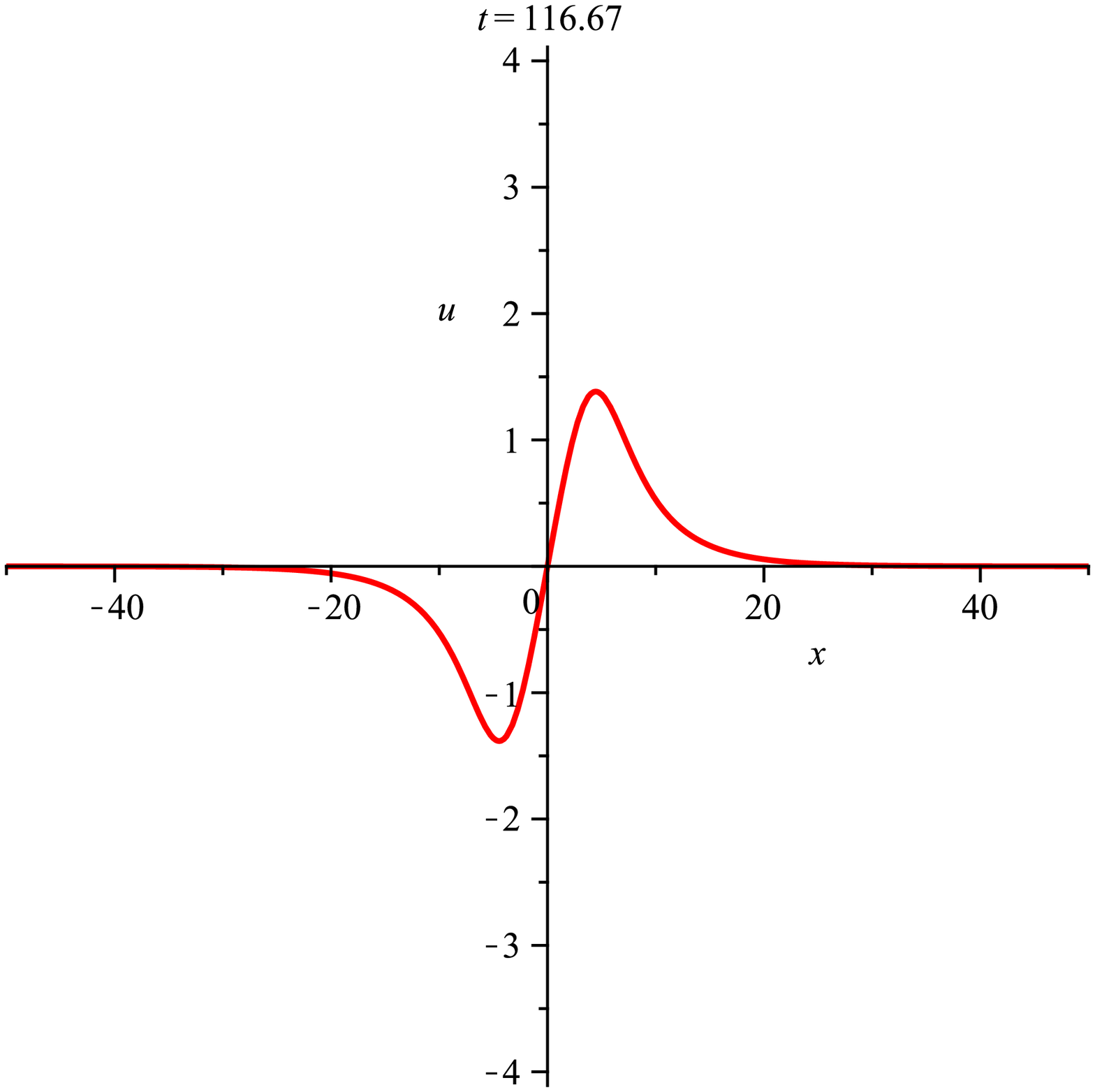} &
\includegraphics[width=0.3\textwidth]{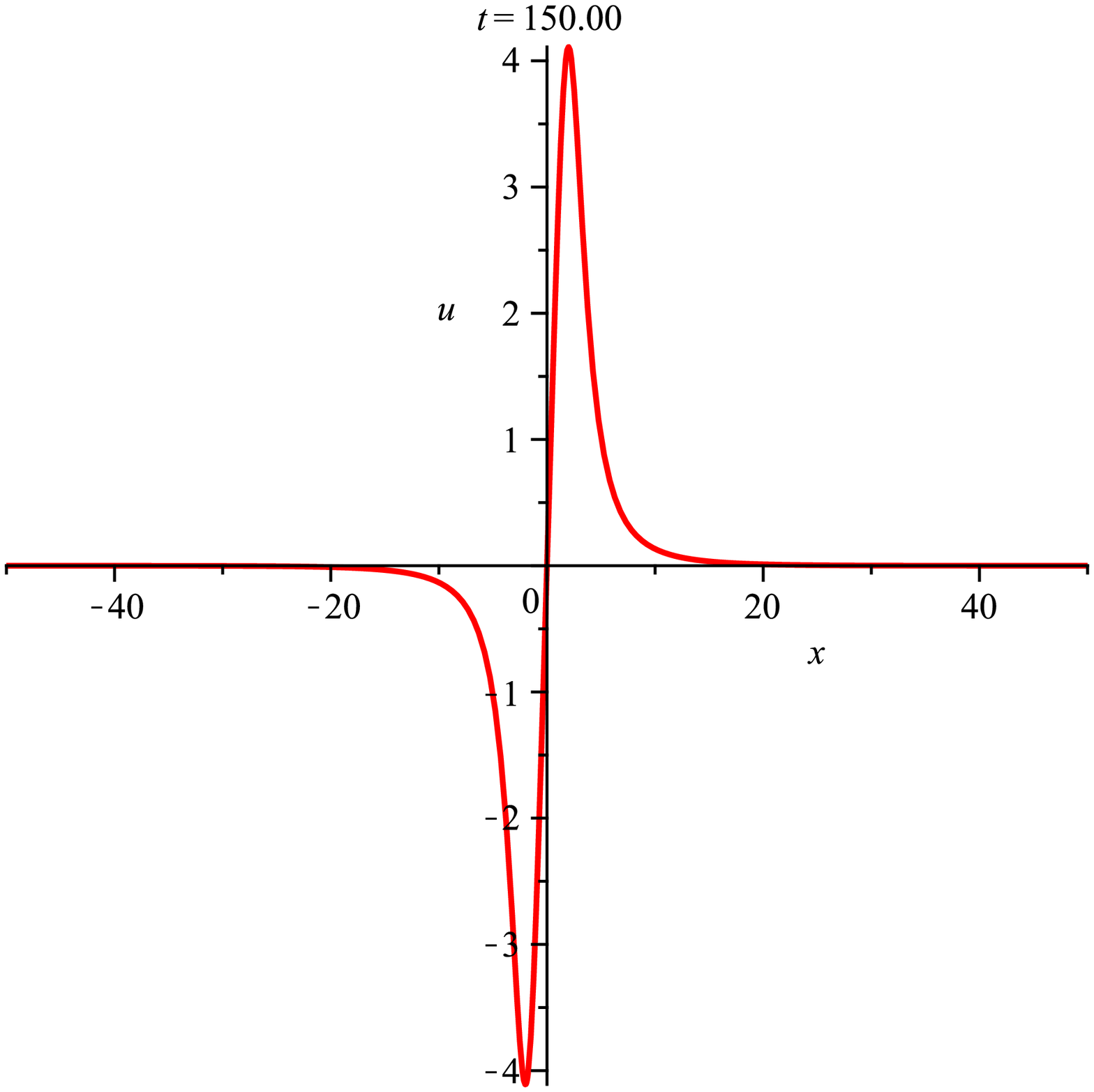} &
\includegraphics[width=0.3\textwidth]{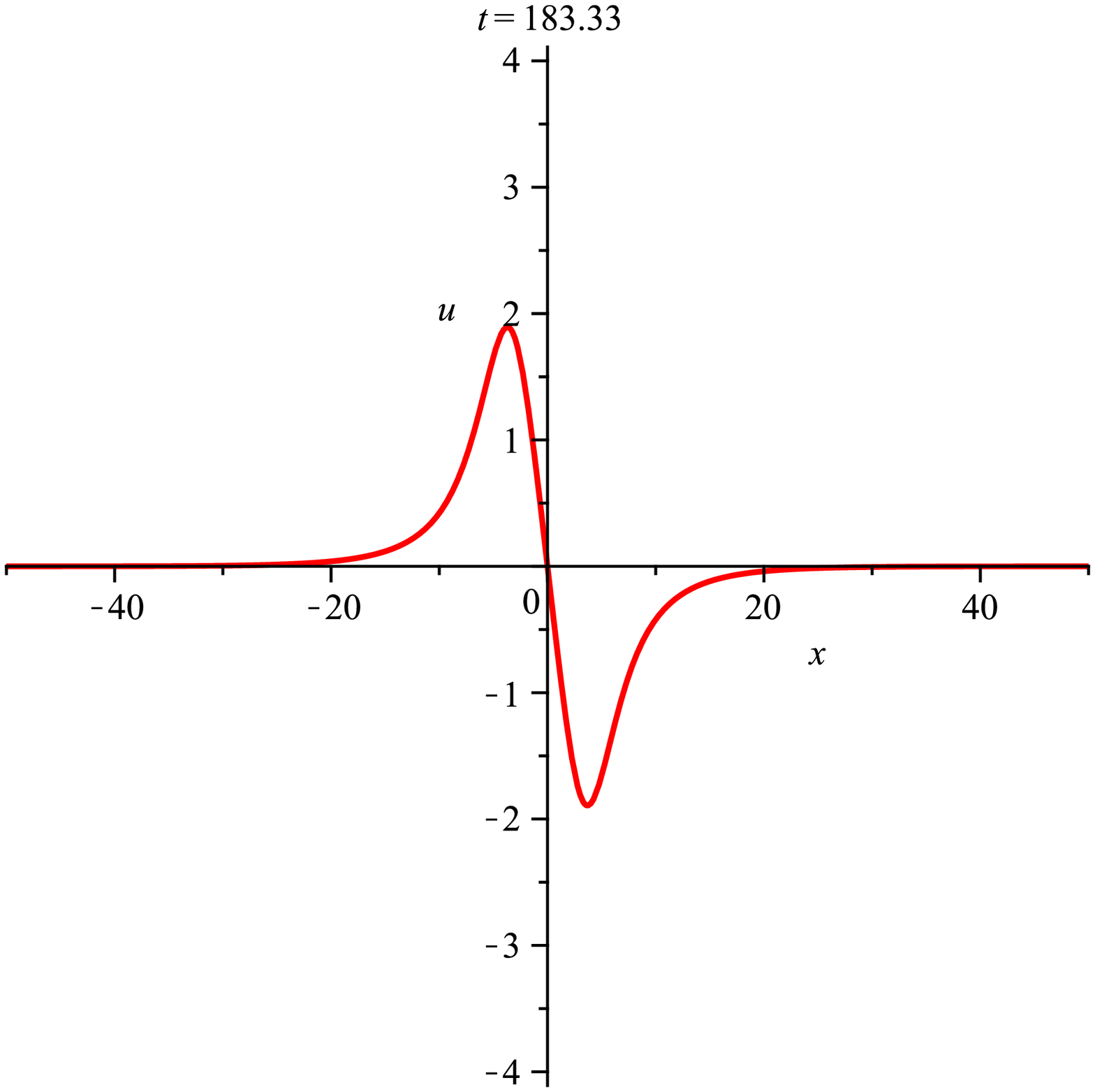} \\
\includegraphics[width=0.3\textwidth]{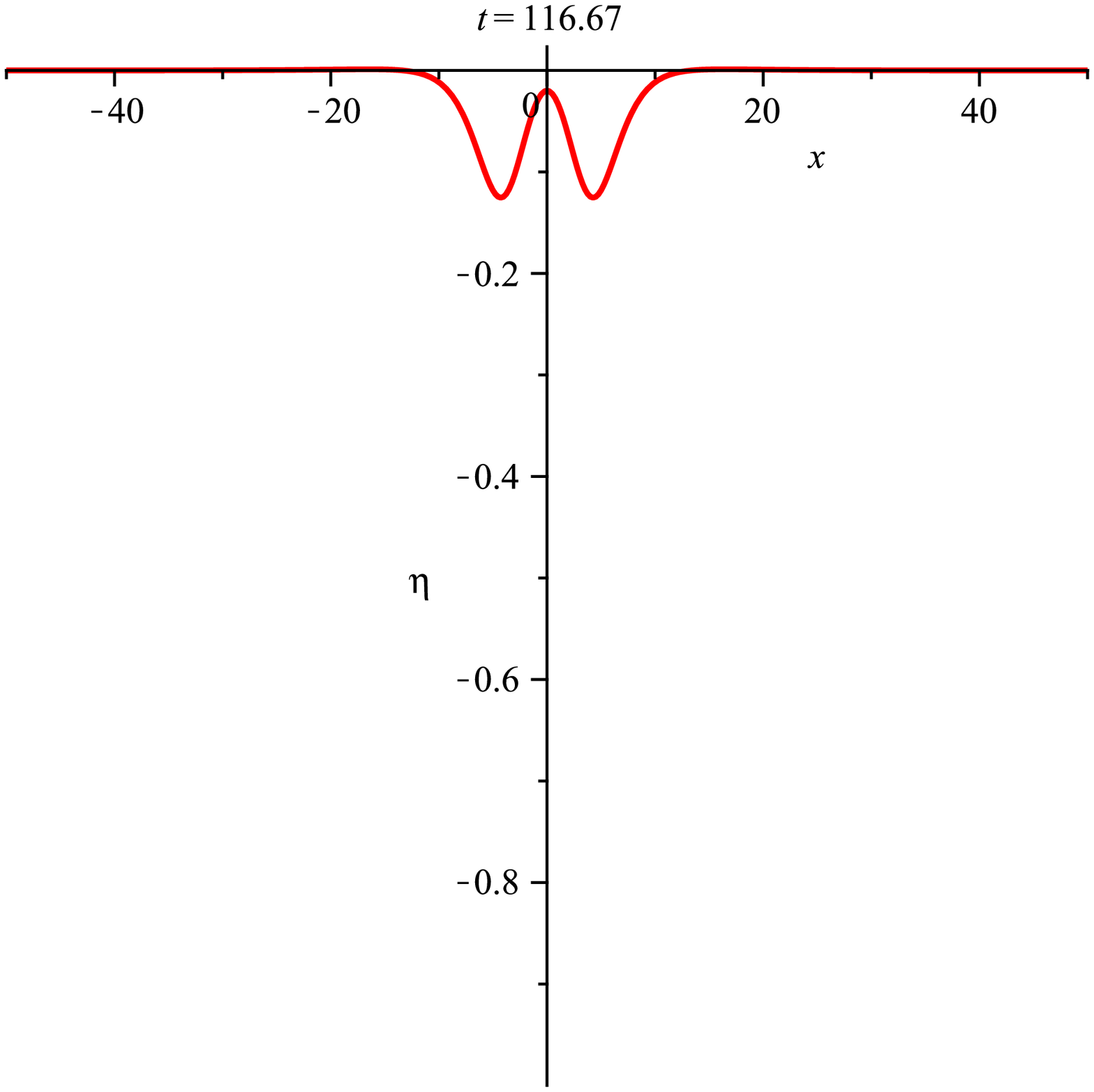} &
\includegraphics[width=0.3\textwidth]{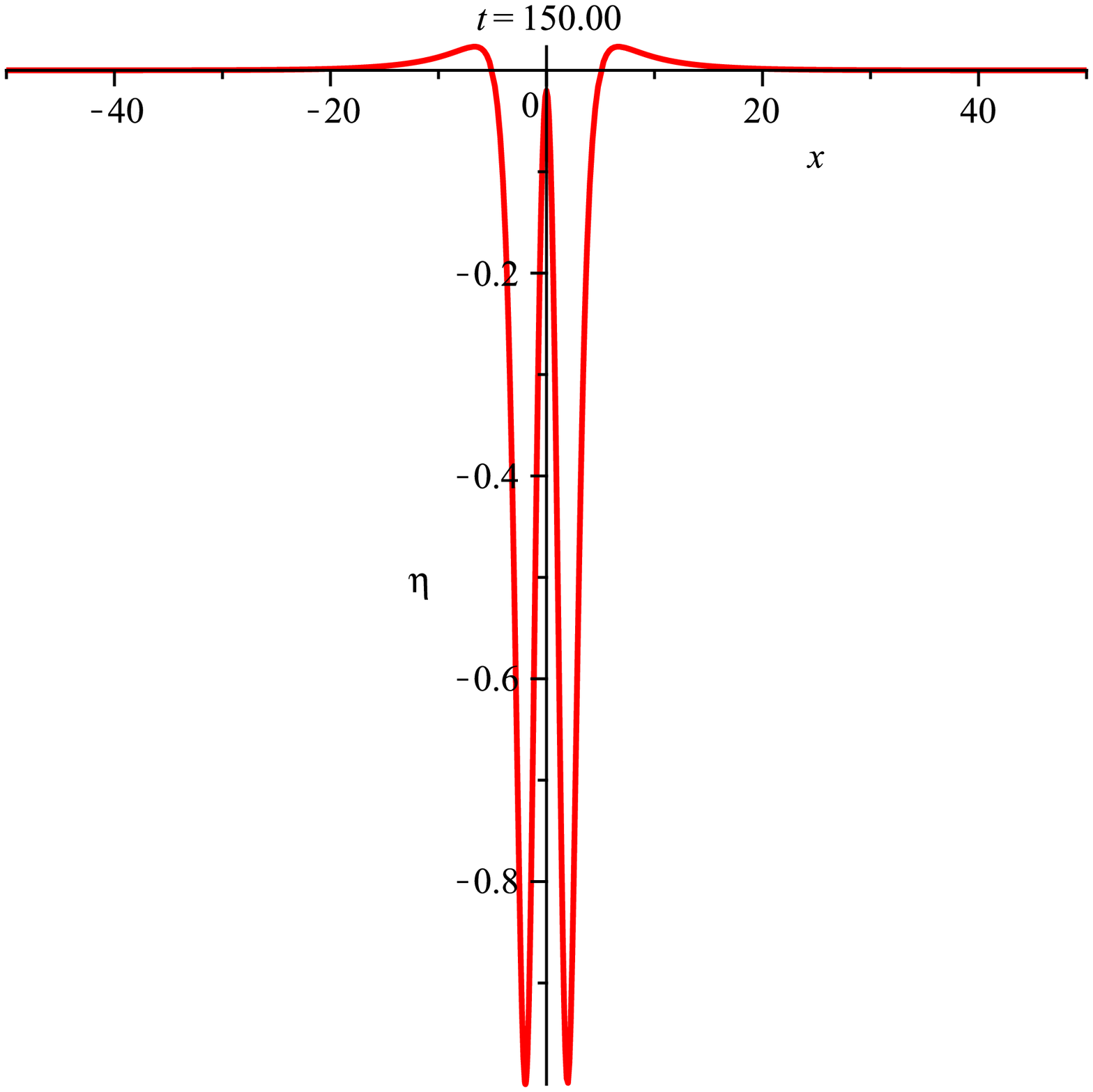} &
\includegraphics[width=0.3\textwidth]{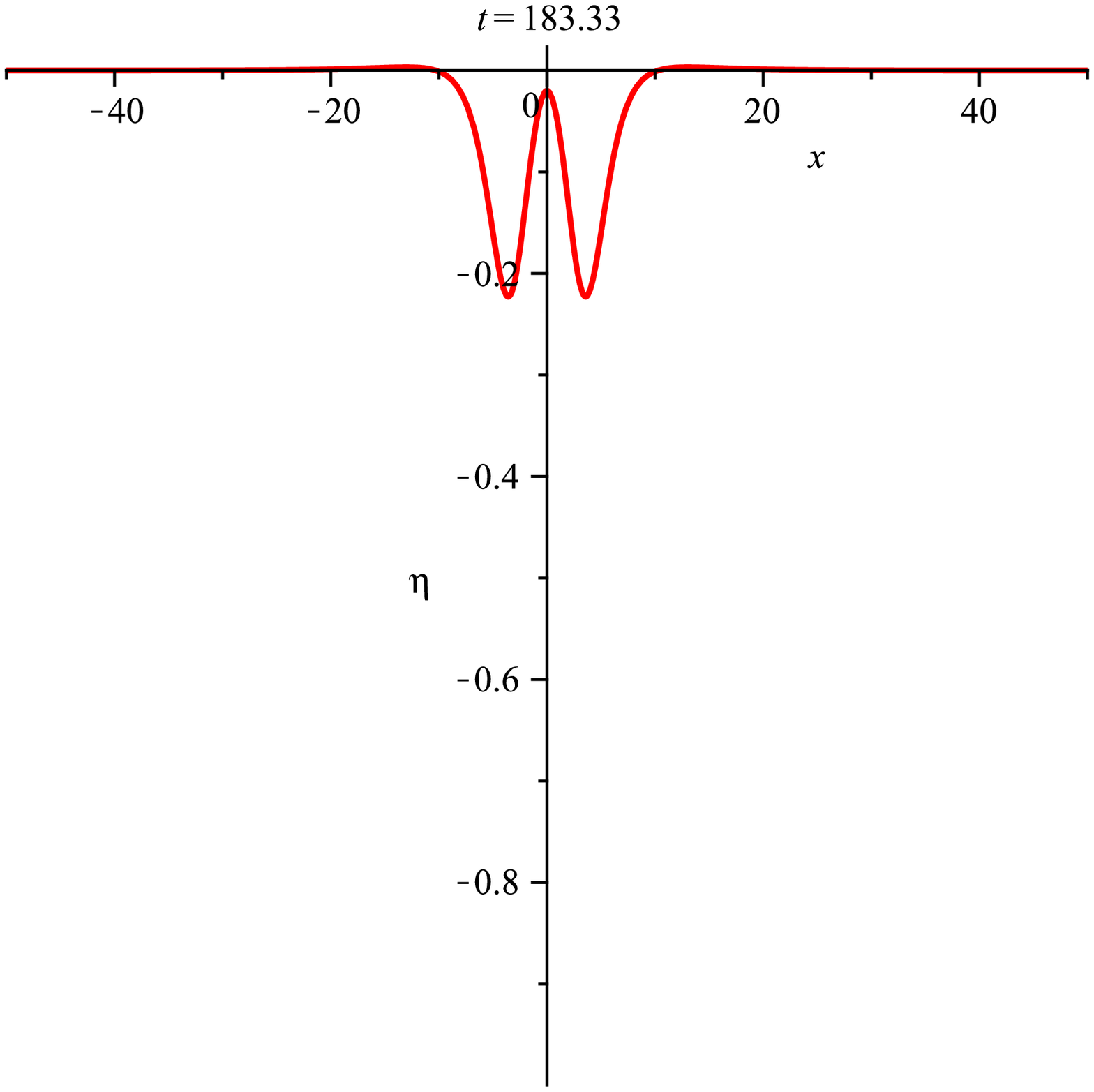}
\end{tabular}
\caption{Snapshots of the solutions of the KB equation (\ref{u}),
(\ref{eta}) for three values of $t$. The first panel is before,
the third panel is after the blowup. } \label{fig2}
\end{figure}

The solution (\ref{u}), (\ref{eta}) is presented on Fig.
\ref{fig2}. Note that $u$ is an odd and $\eta$ is an even function
of $x$. The solution is of 'breather' type and develops
singularities 'infinitely' close to $x=0$ at countably many
isolated values of $t$.

The next case is a solution with $N=2$ discrete eigenvalues. Due
to (\ref{Z2}) there are the following situations:

(i) Both eigenvalues are on the imaginary axis:
$\lambda_1=i\nu_1$, $\lambda_2=i\nu_2$ for some real and positive
$\nu_{1}$ and $\nu_2$;

(ii) $\lambda_2= - \bar{\lambda}_1$,
$R_2(\sigma)=\bar{R}_1(-\sigma)$.  For the (ii) case from
(\ref{Riemann-Hilbert}) we have

\begin{eqnarray}
\underline{\psi}(x;\lambda,\sigma) \!\!&=& \!\!1 \!+\!
ie^{2i\sigma\omega_{+}}\left[\frac{R_{1}(\sigma)e^{2i\lambda_{1}x}\underline{\bar{\psi}}
(x,\bar{\lambda}_{1},\sigma)}{\lambda-\lambda_{1}
}\!+\!\frac{\bar{R}_{1}(\!-\!\sigma)e^{\!-\!2i\bar{\lambda}_{1}x}\underline{\psi}
(x,\bar{\lambda}_{1},\!-\!\sigma)}{\lambda +
\bar{\lambda}_1}\right],\nonumber \\
&\phantom{*}& \label{scalar_2}
\end{eqnarray}

From (\ref{scalar_2}) we obtain a linear system of four equations
for the quantities $\underline{\psi} (x,\bar{\lambda}_{1},\pm
\sigma)$ and their complex conjugates by writing (\ref{scalar_2})
for $\lambda=\bar{\lambda}_1$, the same with $\sigma$ replaced by
$-\sigma$ and their complex conjugates.

The case with $N>2$ eigenvalues is always a combination between
(i) and (ii)  - in general it involves eigenvalues on the
imaginary avis as well as conjugate couples $\lambda_k$ and
$-\bar{\lambda}_k$.

\section{Conclusions} We have outlined the inverse scattering for
the spectral problems of the form (\ref{Lax.2}) with real
functions in the potential, which necesitates the $\mathbb{Z}_2$
reduction (\ref{Z2}). The soliton solution in the case of a single
pole of the eigenfunction does not have the form of a travelling
wave and develops singularities with time. This solution is
probably not relevant for the theory of water waves. There is
another feature of this type of equations which points in the
direction that the purely soliton solutions are probably not the
ones which are observed in the context of water waves. Indeed,
since $\eta$ is the deviation from the equilibrium surface, then
one expects that its space-average value is zero,
$\int_{-\infty}^{\infty} \eta(x,t)\text{d} x=0$. However, the
trace identities which can be derived easily (see e.g.
\cite{Lap07}) for the $N$-soliton solution of the KB equation lead
to the following result:

$$\int_{-\infty}^{\infty} \eta(x,t)\text{d} x =
\frac{1}{4}\int_{-\infty}^{\infty} (u^2+4w)\text{d}
x=-4\sum_{k=1}^{N}\text{Im} \lambda_k.$$

By assumption $\text{Im} \lambda_k >0$ since $\lambda_k$ are in
the upper half complex plane. Thus, we have the following 'mostly
negative' result for the $N$-soliton solution:
$$\int_{-\infty}^{\infty} \eta(x,t)\text{d} x<0.$$

This results indicates that the water wave solutions are related
only to the continuous spectrum and are therefore unstable. This
agrees with the fact that the travelling wave solutions to the
Euler's equation with zero surface tension are unstable.

\section{Acknowledgments}

The authors are indebted to Prof. V.S. Gerdjikov for many valuable
discussions. This material is based upon works supported by the
Science Foundation Ireland (SFI), under Grant No. 09/RFP/MTH2144.

\end{document}